\documentclass{mem}
\usepackage{natbib}
\usepackage{txfonts}
\usepackage{balance}
\usepackage{flushend}
\usepackage{graphicx}
\usepackage{xcolor}
\usepackage[breaklinks,pdftex]{hyperref}
\usepackage[normalem]{ulem}

\begin{document}

\title{
Simulating the quasi-ballistic regime of a short Gamma-Ray Burst jet
}

\author{
E. \,Dreas\inst{1,2,3} 
\and A. \,Pavan\inst{2,4} \and \,R. Ciolfi\inst{2,4} \and A. Celotti\inst{1,3,5}
          }

\institute{SISSA, Via Bonomea 265, I-34136 Trieste, Italy            
        \and
             INAF, Osservatorio Astronomico di Padova, Vicolo dell'Osservatorio 5, I-35122 Padova, Italy
        \and
        INFN, Sezione di Trieste, Via Valerio 2, I-34127 Trieste, Italy
        \and
        INFN, Sezione di Padova, Via Francesco Marzolo 8, I-35131 Padova, Italy
        \and
        INAF, Osservatorio Astronomico di Brera, via Bianchi 46, I-23807 Merate, Italy\\ \email{edreas@sissa.it}\\
}


\titlerunning{Quasi-ballistic sGRB jets}

\date{Received: XX-XX-XXXX (Day-Month-Year); Accepted: XX-XX-XXXX (Day-Month-Year)}

\abstract{
This study extends the 3D magnetohydrodynamic (MHD) simulation of a jet emerging from a binary neutron star (BNS) merger presented in Pavan et al.~(2023), in which an incipient jet was manually injected into the realistic environment imported from a previous general-relativistic MHD simulation of a merging BNS system. The jet evolution is followed up to almost 10 seconds without loss of resolution. Our results reveal that the jet faces challenges in penetrating the dense surroundings, leading to a barely successful outflow that exhibits structural asymmetries and low Lorentz factors. By the end of the extended simulation, 98\% of the jet energy is converted to kinetic form and its angular structure is stabilized. The physical quantities inferred thus provide reliable inputs for afterglow emission calculations. This work demonstrates a method for simulating jets in 3D up to nearly ballistic regimes that is general and ready to be applied to any jet in a BNS merger context.

}
\maketitle{}

\section{Introduction}

Gamma-ray bursts (GRBs) are explosive phenomena characterized by an early sub-MeV prompt emission phase, followed by a prolonged afterglow phase spanning from radio to TeV wavelengths. Both phases originate from the dissipation of energy in collimated relativistic jets powered by compact central engines, such as hyper-accreting black holes or magnetized neutron stars \citep[e.g.][]{Kumar2015}. The afterglow is produced when these jets interact with the surrounding medium, producing shocks and radiative dissipation via the synchrotron and inverse-Compton processes. 
GRBs are expected to originate from either collapsed massive stars or compact binary mergers \citep[e.g.][]{Zhang2018}.
The association between short GRBs and binary neutron star (BNS) mergers was confirmed by the coincidence of GRB\,170817A with the gravitational wave signal GW170817 associated with a BNS merger \citep{LVC-BNS,LVC-MMA,LVC-GRB}. 

The angular structure of these jets, shaped during their launch and propagation through post-merger environments, freezes while approaching the ballistic phase, i.e.~when the energy is almost entirely converted into kinetic form. The outflow then expands freely until interactions with the interstellar medium give rise to the afterglow radiation \citep[][and refs.~therein]{Salafia2022}.
Therefore, the afterglow signals contain information about the jet launching conditions and the propagation environment, complementing gravitational wave data. 

Relativistic MHD simulations \citep[e.g.][]{Lazzati2017,Murguia2021,Geng2019,Urrutia2023,Gottlieb2023a}—recently also accounting for realistic post-merger environments \citep{Pavan2021,Pavan2023,Lazzati2021}
—enable modelling of the jet propagation and its angular structure. While 2D simulations often reach the ballistic regime needed for accurate afterglow light curve predictions \citep{Lazzati2018,Xie2018,Urrutia2021}, 3D simulations typically do not last sufficiently long to get close to this regime
(e.g.~\citealt{Kathirgamaraju2019,Nathanail2021,Nativi2022}). 

Here, we focus on a specific short GRB jet simulation with the goal of approaching such a condition in 3D. We analyse the dynamics, energetics, and structure of the jet throughout its evolution.
What presented below elaborates on the work done in \cite{Dreas2025} (hereafter D25), which we refer to for further details and discussion.

\section{Methods and main results}

As a starting point, we rely on a reference simulation of a short GRB jet employing as initial condition a post-merger environment directly imported, in terms of hydrodynamical properties and magnetic field, from the general relativistic magneto hydrodynamic (GRMHD) simulation presented in \cite{Ciolfi2020b}. In this environment, a magnetized axisymmetric jet is manually injected with the following parameters: half-opening angle $\Theta_\mathrm{j} = 10^\circ$, initial Lorentz factor $\Gamma_\mathrm{j} = 3$, terminal Lorentz factor $\Gamma_{\infty} = 300$, and initial two-sided luminosity of $L_{j} = 3 \times 10^{51}$ erg/s that fades away exponentially on a characteristic timescale of 0.3 s.
The magnetic field contribution to the total injected luminosity is 8\%.
This jet simulation is performed with the PLUTO code \citep{Mignone2007-PLUTO1} and follows the first two seconds of the jet evolution in a 3D spherical grid with logarithmic radial spacing and $r_\mathrm{max}$ = $2.5 \times 10^6$; we refer to \citealt{Pavan2023} (P23 henceforth) for details. 

With the aim of producing a more reliable input for calculating the light curves of the afterglow emission produced at much later times (days to years after merger), we extend the simulation presented above up to almost 10 s, focussing on the north-side only.
To do so, we first continue the same simulation for another second (i.e.~up to 3\,s), and then we import the north-side jet onto a Cartesian domain, allowing for the computational cells to be uniform, in our case cubical, and thus avoiding the loss of resolution at increasing radial scales intrinsic to the spherical grid.
In particular, the cell spacing in the Cartesian grid is of 6800\,km, corresponding to a resolution that, in the outer high-velocity regions of the outflow (fluid elements with $\Gamma \gtrsim 2 $), is higher compared to the original spherical grid.
To balance resolution and computational efficiency, we use $Nx = Nz = 848$ points along $x$ and $z$ and $Ny = 512$ points along $y$ (the jet injection axis), spanning $[-2.88, 2.88]\times 10^6$\,km in $x,z$, and $[0.1,3.58] \times 10^6$\,km in $y$. 
Moreover, we quantitatively define the radial range that contains most of the jet's energy, that we refer to as the `jet head', as the region where the radial flux of energy is above 15\% of the overall maximum at each given time. Such a jet head definition is used for our analysis of the jet structure.
\begin{figure*}[t!]
\resizebox{\hsize}{!}{\includegraphics[clip=true]{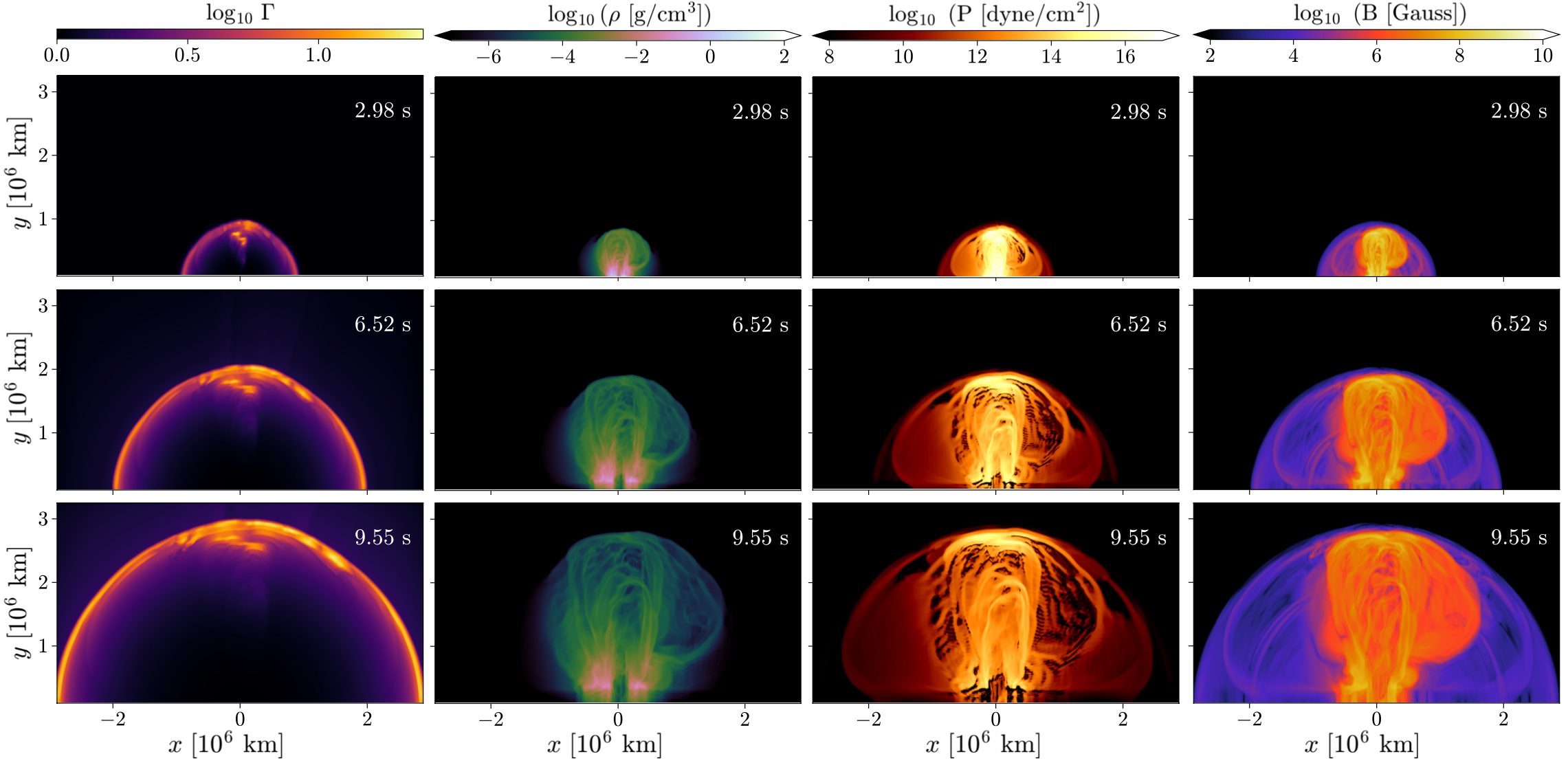}}
\caption{\footnotesize
Meridional view of Lorentz factor, rest-mass density, total pressure, and magnetic field strength (left to right) of the simulation at three different times after jet injection (top to bottom): $ \approx\!3$\,s (beginning of the evolution on the new Cartesian grid), $\approx\!6.5$\,s, and $\approx\!9.5$\,s.}
    \label{fig:cart_panel}
\end{figure*}
\begin{figure}
\resizebox{\hsize}{!}{\includegraphics[clip=true]{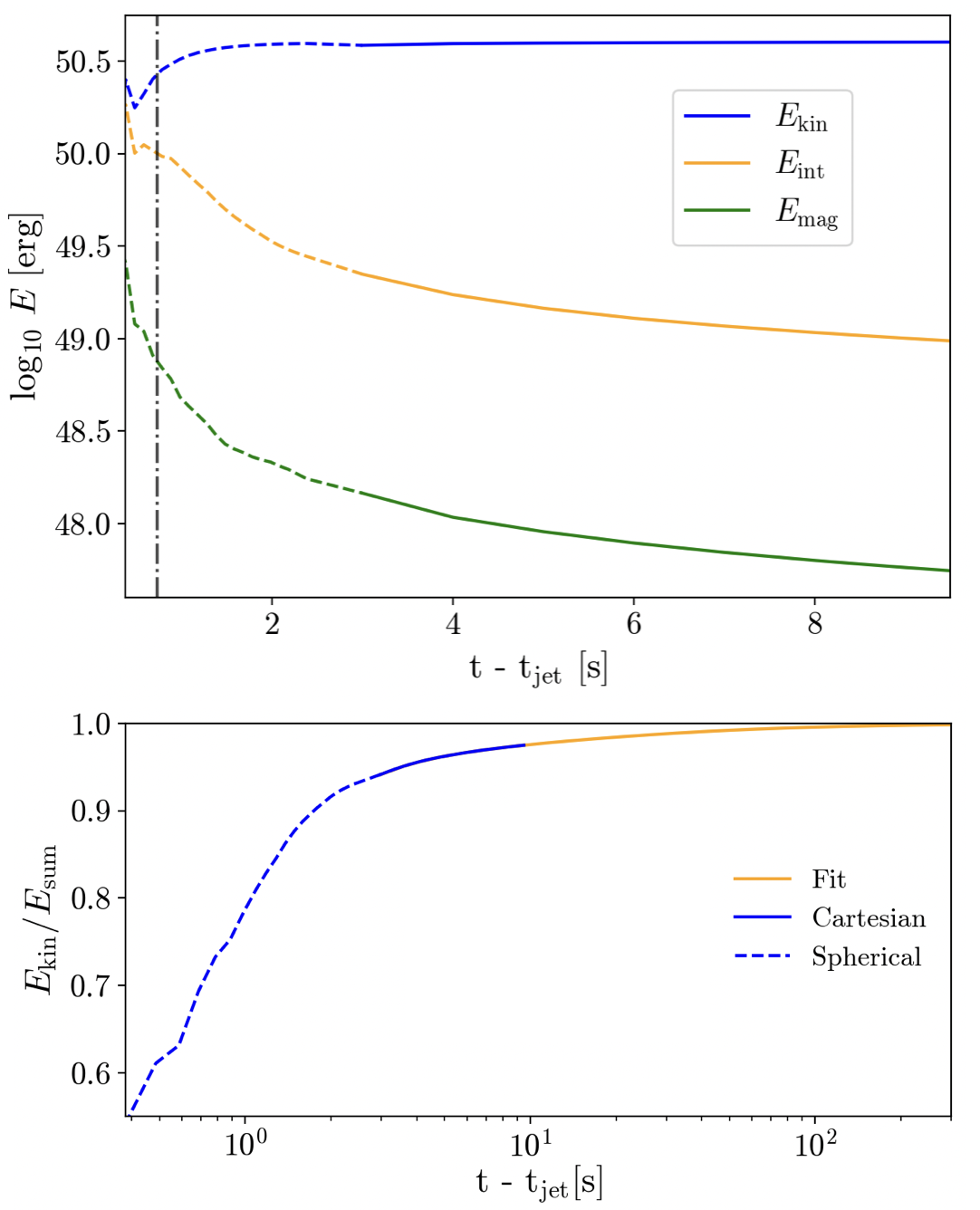}}
\caption{
\footnotesize
Upper panel: time evolution of the different energy components in the whole computational domain: kinetic ($E_\mathrm{kin}$), internal ($E_\mathrm{int}$) and magnetic ($E_\mathrm{mag}$). The dashed (continuous) lines refer to the evolution employing the original spherical (Cartesian) grids. The vertical line represents the breakout time (see text). Lower panel: evolution of the fraction of $E_\mathrm{kin}/E_\mathrm{sum}$ (see text) within the simulation time in the spherical and Cartesian grids (in blue), further extrapolated to later times via a polynomial fit (orange line). Time is represented in logarithmic scale.
}
\label{fig:ecomp}
\end{figure}

    \subsection{Dynamics and energetics}\label{sec:rr_obs}

In the first second of evolution the interplay between the jet and the post-merger environment strongly affects the evolution. In fact, the jet spends a consistent part of its energy (about 50 \%) to break out from the dense surrounding material, leading to final Lorentz factors considerably smaller than the analytical upper limit set at injection ($\Gamma_\infty=300$). Moreover, the asymmetric distribution of the `realistic' environment affects the evolution: the jet will preferably travel trough lower density regions, and that will in turn lead to an angular jet structure that differs from typical top-hat or gaussian models (see below).
In this simulation, as presented in P23, the jet `breaks out' of the surrounding material at $t_\mathrm{BO} \simeq 350$ ms after jet launching ($t_\mathrm{jet}$) at a distance of $r_\mathrm{BO} \simeq 5 \times 10^4$ km.

The results of the `extended' Cartesian evolution (i.e.~from 3 s on), in terms of Lorentz factor, mass density, pressure, and magnetic field intensity are presented in Fig. \ref{fig:cart_panel} at the beginning of the simulation ($ \approx$ 3 s), at an intermediate time ($\approx$ 6.5 s), and at the end of the simulation ($\approx$ 9.5 s).
We observe (i) a residual acceleration of the fastest part of the jet, due to further conversion of internal and magnetic energetic components into kinetic form (see below); (ii) an evolution of the outflow from a more elongated, jet-like structure, to a portion of a spherical shell.

The outflow that we observe consists of mixed jet-environment material exhibiting a complex morphology shaped by the previous interactions (see the distributions of all the quantities presented in the Figure). 
\begin{figure*}[t!]
\resizebox{\hsize}{!}{\includegraphics[clip=true]{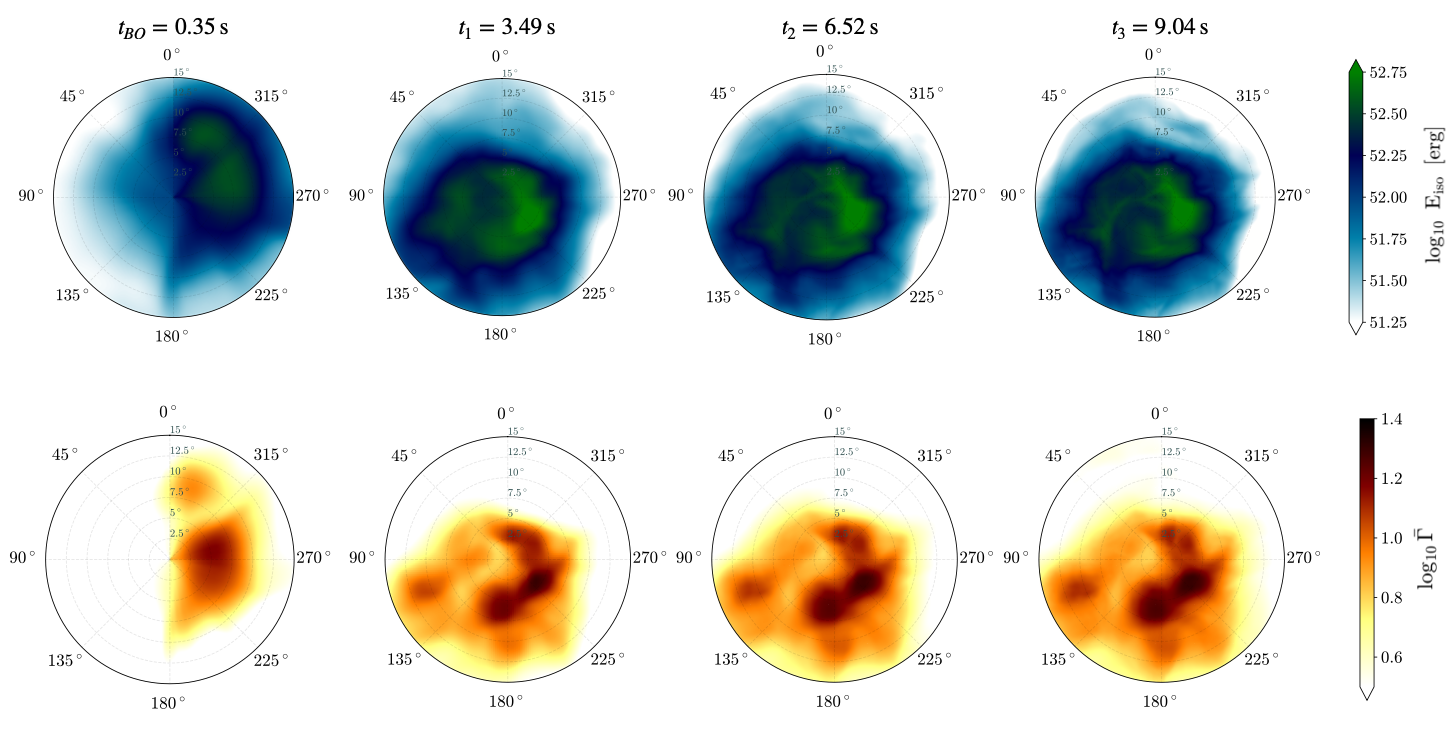}}
\caption{\footnotesize
2D front-view angular distribution of the isotropic equivalent energy (top) and the radially averaged Lorentz factor (bottom) of the jet head at the times (after jet launching) 
$t_{BO}$, $t_1$, $t_2$ and $t_3$ specified at the top}. The jet injection axis corresponds to the central point in each panel.
    
    \label{fig:egm}
\end{figure*}

In Fig.~\ref{fig:ecomp}, we analyse the time evolution of the energetic budget in the whole computational domain, from $t_\mathrm{jet}$ to the end of the extended evolution.
Indeed, we observe an overall continuous conversion of the internal and magnetic components into the kinetic one after the break-out, while the evolution in the first few milliseconds is affected by many complex processes (see P23 and D25). 
The lower panel of the Figure shows the fraction of the total energy without rest-mass component ($E_\mathrm{sum}$) that is in kinetic form as a function of time, which reaches about 98\% by the end of the extended simulation.

\subsection{Angular distributions}
Another characteristic of approaching the ballistic regime is the freezing of the jet angular structure. In Fig.~\ref{fig:egm} we show (at three different times) the front-view distributions of two quantities relevant for the following radiative emission: the isotropic equivalent energy of the jet head 
\begin{equation}    
    E_{\mathrm{iso}}(\Theta,\Phi) =  4 \pi \int_{r_\mathrm{in}}^{r_\mathrm{out}} r^2 e_{\mathrm{sum}}(r,\Theta,\Phi) dr \,\, ,
\end{equation}
where $e_{\mathrm{sum}}$ is the sum of kinetic, internal and magnetic energy densities;
and the radially averaged Lorentz factor weighted on $e_{\mathrm{sum}}$
\begin{equation}
    \bar{\Gamma}(\Theta,\Phi) = \frac{\int_{r_\mathrm{in}}^{r_\mathrm{out}} \Gamma(r,\Theta,\Phi) e_{\mathrm{sum}}(r,\Theta,\Phi) r^2  dr}{\int_{r_\mathrm{in}}^{r_\mathrm{out}}  e_{\mathrm{sum}}(r,\Theta,\Phi) r^2 dr}  \,\, .
\label{gammabar}
\end{equation}

As visible in the Figure, there are significant modifications in their distribution from $t_\mathrm{BO}$ to 3 seconds. Some residual changes in terms of acceleration and structure are observed also during the extended simulation. This evolution is however strongly reduced towards the end of the simulation, indicating a converging trend of the angular structure (we refer to 5 for a detailed discussion). 

Fig.~\ref{fig:egm} also illustrates how the distribution of $E_\mathrm{iso}$ and $\bar{\Gamma}$ deviates from axisymmetry. 
In particular, we note that local maxima in both distributions are observed at about $4^\circ$ from the injection axis. 

\section{Future perspectives}
 The specific jet considered reaches only a modest average Lorentz factor of $\sim$ 10, too low to produce a canonical sGRB, but still representing a possible physical case.
A new set of simulations with different injection parameters, including cases corresponding to typical sGRB jets, is under development. The approach followed in this work will be applied systematically to all of the new simulations. We further plan to compute the multi-wavelength afterglow light curves for all cases, allowing us to investigate how different jet properties shape the observable electromagnetic emission.



\begin{acknowledgements}
We thank Om Sharan Salafia for insightful discussions. We further thank Giancarlo Ghirlanda and Andrea Mignone for their useful comments.
This work was supported by the European Union under NextGenerationEU, via the PRIN 2022 Projects ``EMERGE: Neutron star mergers and the origin of short gamma-ray bursts", Prot. n. 2022KX2Z3B (CUP C53D23001150006), and ``PEACE: Powerful Emission and Acceleration in the most powerful Cosmic Explosion'', Prot. n. 202298J7KT (CUP G53D23000880006). 
The views and opinions expressed are solely those of the authors and do not necessarily reflect those of the European Union, nor can the European Union be held responsible for them.
AP and RC acknowledge further support by the Italian Ministry of Foreign Affairs and International Cooperation (MAECI), grant number US23GR08.  
Simulations were performed on the Discoverer HPC cluster at Sofia Tech Park (Bulgaria). We acknowledge EuroHPC Joint Undertaking for awarding us access to this cluster via the Regular Access allocations EHPC-REG-2022R03-218 and EHPC-REG-2023R03-160.
\end{acknowledgements}

\bibpunct{(}{)}{;}{a}{}{,} 
\bibliographystyle{aa} 
\bibliography{refs} 

\begin{thebibliography}{23}
\expandafter\ifx\csname natexlab\endcsname\relax\def\natexlab#1{#1}\fi

\bibitem[{{Abbott} {et~al.}(2017{\natexlab{a}}){Abbott}, {Abbott}, {Abbott}, {Acernese}, {Ackley}, {Adams}, {Adams}, {Addesso}, {Adhikari}, {Adya}, \& et~al.}]{LVC-GRB}
{Abbott}, B.~P., {Abbott}, R., {Abbott}, T.~D., {et~al.} 2017{\natexlab{a}}, \apjl, 848, L13

\bibitem[{{Abbott} {et~al.}(2017{\natexlab{b}}){Abbott}, {Abbott}, {Abbott}, {Acernese}, {Ackley}, {Adams}, {Adams}, {Addesso}, {Adhikari}, {Adya}, \& et~al.}]{LVC-BNS}
{Abbott}, B.~P., {Abbott}, R., {Abbott}, T.~D., {et~al.} 2017{\natexlab{b}}, \prl, 119, 161101

\bibitem[{{Abbott} {et~al.}(2017{\natexlab{c}}){Abbott}, {Abbott}, {Abbott}, {Acernese}, {Ackley}, {Adams}, {Adams}, {Addesso}, {Adhikari}, {Adya}, \& et~al.}]{LVC-MMA}
{Abbott}, B.~P., {Abbott}, R., {Abbott}, T.~D., {et~al.} 2017{\natexlab{c}}, \apjl, 848, L12

\bibitem[{{Ciolfi}(2020)}]{Ciolfi2020b}
{Ciolfi}, R. 2020, Gen. Rel. Grav., 52, 59

\bibitem[{{Dreas} {et~al.}(2025){Dreas}, {Pavan}, {Ciolfi}, \& {Celotti}}]{Dreas2025}
{Dreas}, E., {Pavan}, A., {Ciolfi}, R., \& {Celotti}, A. 2025, \aap, 694, A200

\bibitem[{{Geng} {et~al.}(2019){Geng}, {Zhang}, {K{\"o}lligan}, {Kuiper}, \& {Huang}}]{Geng2019}
{Geng}, J.-J., {Zhang}, B., {K{\"o}lligan}, A., {Kuiper}, R., \& {Huang}, Y.-F. 2019, \apjl, 877, L40

\bibitem[{Gottlieb {et~al.}(2023)Gottlieb, Metzger, Quataert, Issa, Martineau, Foucart, Duez, Kidder, Pfeiffer, \& Scheel}]{Gottlieb2023a}
Gottlieb, O., Metzger, B.~D., Quataert, E., {et~al.} 2023, \apjl, 958

\bibitem[{{Kathirgamaraju} {et~al.}(2019){Kathirgamaraju}, {Tchekhovskoy}, {Giannios}, \& {Barniol Duran}}]{Kathirgamaraju2019}
{Kathirgamaraju}, A., {Tchekhovskoy}, A., {Giannios}, D., \& {Barniol Duran}, R. 2019, \mnras, 484, L98

\bibitem[{{Kumar} \& {Zhang}(2015)}]{Kumar2015}
{Kumar}, P. \& {Zhang}, B. 2015, \physrep, 561, 1

\bibitem[{{Lazzati} {et~al.}(2017){Lazzati}, {L{\'o}pez-C{\'a}mara}, {Cantiello}, {Morsony}, {Perna}, \& {Workman}}]{Lazzati2017}
{Lazzati}, D., {L{\'o}pez-C{\'a}mara}, D., {Cantiello}, M., {et~al.} 2017, \apjl, 848, L6

\bibitem[{{Lazzati} {et~al.}(2021){Lazzati}, {Perna}, {Ciolfi}, {Giacomazzo}, {L{\'o}pez-C{\'a}mara}, \& {Morsony}}]{Lazzati2021}
{Lazzati}, D., {Perna}, R., {Ciolfi}, R., {et~al.} 2021, \apjl, 918, L6

\bibitem[{{Lazzati} {et~al.}(2018){Lazzati}, {Perna}, {Morsony}, {Lopez-Camara}, {Cantiello}, {Ciolfi}, {Giacomazzo}, \& {Workman}}]{Lazzati2018}
{Lazzati}, D., {Perna}, R., {Morsony}, B.~J., {et~al.} 2018, \prl, 120, 241103

\bibitem[{{Mignone} {et~al.}(2007){Mignone}, {Bodo}, {Massaglia}, {Matsakos}, {Tesileanu}, {Zanni}, \& {Ferrari}}]{Mignone2007-PLUTO1}
{Mignone}, A., {Bodo}, G., {Massaglia}, S., {et~al.} 2007, \apjs, 170, 228

\bibitem[{{Murguia-Berthier} {et~al.}(2021){Murguia-Berthier}, {Ramirez-Ruiz}, {De Colle}, {Janiuk}, {Rosswog}, \& {Lee}}]{Murguia2021}
{Murguia-Berthier}, A., {Ramirez-Ruiz}, E., {De Colle}, F., {et~al.} 2021, \apj, 908, 152

\bibitem[{{Nathanail} {et~al.}(2021){Nathanail}, {Gill}, {Porth}, {Fromm}, \& {Rezzolla}}]{Nathanail2021}
{Nathanail}, A., {Gill}, R., {Porth}, O., {Fromm}, C.~M., \& {Rezzolla}, L. 2021, \mnras, 502, 1843

\bibitem[{{Nativi} {et~al.}(2022){Nativi}, {Lamb}, {Rosswog}, {Lundman}, \& {Kowal}}]{Nativi2022}
{Nativi}, L., {Lamb}, G.~P., {Rosswog}, S., {Lundman}, C., \& {Kowal}, G. 2022, \mnras, 509, 903

\bibitem[{{Pavan} {et~al.}(2021){Pavan}, {Ciolfi}, {Kalinani}, \& {Mignone}}]{Pavan2021}
{Pavan}, A., {Ciolfi}, R., {Kalinani}, J.~V., \& {Mignone}, A. 2021, \mnras, 506, 3483

\bibitem[{{Pavan} {et~al.}(2023){Pavan}, {Ciolfi}, {Kalinani}, \& {Mignone}}]{Pavan2023}
{Pavan}, A., {Ciolfi}, R., {Kalinani}, J.~V., \& {Mignone}, A. 2023, \mnras, 524, 260

\bibitem[{{Salafia} \& {Ghirlanda}(2022)}]{Salafia2022}
{Salafia}, O.~S. \& {Ghirlanda}, G. 2022, Galaxies, 10, 93

\bibitem[{{Urrutia} {et~al.}(2023){Urrutia}, {De Colle}, \& {L{\'o}pez-C{\'a}mara}}]{Urrutia2023}
{Urrutia}, G., {De Colle}, F., \& {L{\'o}pez-C{\'a}mara}, D. 2023, \mnras, 518, 5145

\bibitem[{{Urrutia} {et~al.}(2021){Urrutia}, {De Colle}, {Murguia-Berthier}, \& {Ramirez-Ruiz}}]{Urrutia2021}
{Urrutia}, G., {De Colle}, F., {Murguia-Berthier}, A., \& {Ramirez-Ruiz}, E. 2021, \mnras, 503, 4363

\bibitem[{{Xie} {et~al.}(2018){Xie}, {Zrake}, \& {MacFadyen}}]{Xie2018}
{Xie}, X., {Zrake}, J., \& {MacFadyen}, A. 2018, \apj, 863, 58

\bibitem[{{Zhang}(2018)}]{Zhang2018}
{Zhang}, B. 2018, {The Physics of Gamma-Ray Bursts} ({Cambridge Univeristy Press, ISBN: 978-1-139-22653-0})

\end{thebibliography}

\end{document}